\documentclass{emulateapj}
\usepackage{natbib}

\newcommand{\eps}{{\rm ergs\,s^{-1}}}
\newcommand{\epcs}{{\rm ergs\,cm^{-2}\,s^{-1}}}
\newcommand{\cts}{{\rm count\,s^{-1}}}

\newcommand{\src}{CH~Cyg}

\slugcomment{Accepted by ApJL, August 9 2004}

\shorttitle{An X-ray jet from CH~Cygni}
\shortauthors{Galloway \& Sokoloski}

\begin{document}

\title{An X-Ray Jet from a White Dwarf -- \\ Detection of the Collimated
Outflow from CH~Cygni with {\it Chandra}}

\author{Duncan K. Galloway}
\affil{ Center for Space Research, 37-626b,
   Massachusetts Institute of Technology, Cambridge, MA 02139}
\email{duncan@space.mit.edu}
\and
\author{J. L. Sokoloski\footnote{NSF Astronomy and Astrophysics Fellow.}}
\affil{Smithsonian Astrophysical Observatory, 60 Garden St.,
Cambridge, MA 02138} 
\email{jsokolos@cfa.harvard.edu}

\begin{abstract}
Most symbiotic stars consist of a white dwarf accreting material from
the wind of a red giant.  An increasing number of these objects have been
found to produce jets.  Analysis of archival {\it Chandra}\/ data
of the symbiotic system CH~Cygni reveals faint extended
emission to the south, aligned with the optical and radio jets seen in
earlier {\it HST}\/ and VLA observations.
CH~Cygni thus contains only the
second known white dwarf with an X-ray jet, after R~Aquarii.
The X-rays from symbiotic-star jets appear to be produced when jet
material is shock-heated following collision with surrounding gas, as with
the outflows from some protostellar objects and bipolar planetary nebulae.
\end{abstract}


\section{Introduction}

Collimated jets are produced by many accreting astrophysical systems,
ranging from proto-stars, through stellar-mass black holes in X-ray
binaries, to active galactic nuclei.  
Symbiotic stars, in which a white dwarf (WD) accretes from the wind of
a red giant, are probably the most recently recognized class of
jet-producing objects.  There are currently at least 10-12 symbiotics
with evidence for bipolar outflows \cite[]{brocksopp04,corradi01}.
Very little is known about the X-ray properties of symbiotic-star
jets, as only one symbiotic jet has been reported in the X-rays
\cite[R~Aquarii; see][]{kellogg01}.

CH~Cygni ($l=81\fdg86$, $b=+15\fdg58$) is one of the best-studied
symbiotic systems \cite[for a review of \src's properties see][]{kenyon01}.  
The multiple photometric and radial velocity periodicities present in
the system suggest the possible presence of a third star
\cite[]{hinkle93}.
For the distance, we adopt $245\pm50$~pc, from the mode of the
Hipparcos parallax probability distribution \cite[]{hipparcos}.
Evidence for a radio jet was first detected between April 1984 and May
1985, coincident with a strong radio outburst and a substantial
decline in the visual magnitude
\cite[]{taylor86}.  Subsequent radio and optical observations 
confirmed the extended emission as arising from a jet
\cite[]{crocker01}, and also revealed that the jet position angle on
the sky changes with time, suggesting that the jet axis precesses with
a period of $\approx6520$~d \cite[]{crocker02}.  Changes in the fast
optical flickering in 1997, after a jet was produced, have also
suggested a direct connection between the accretion disk and jet
production in CH Cyg \cite[]{sk03a}.
X-ray emission was first conclusively detected from CH Cyg by {\it
EXOSAT}\/ on 1985 May 24, at a flux of $1.3\times10^{-11}\ \epcs$
 \cite[0.02--2.5~keV;][]{leahy87}.
Subsequent {\it ASCA}\/ observations revealed a complex X-ray spectrum which
could be fit either with multiple thermal components \cite[]{ezuka98} or a
single-temperature thermal spectrum attenuated by an ionized absorber
\cite[]{wheatley01}.

Here we describe analysis
of a {\it Chandra}\/ observation in which we detect, for the first time,
the X-ray jet in \src.

\section{Observations}

We analyzed an observation of \src\ acquired between 2001
March 27 06:22 and 20:02 UT,
by the {\it Chandra X-ray Observatory}\/ \cite[]{chandra02}.  
The observation 
was made in timed/FAINT mode with the high-energy
transmission grating (HETG) in place, although the source flux was too low
to obtain a high signal-to-noise grating spectrum (the first-order count
rate was $6.3\times10^{-3}\ \cts$). Thus, we did not consider the
dispersed spectra for our analysis.  The CCDs are sensitive to photons in
the 0.2--10~keV energy range, and have a pixel size of $0\farcs492$.  The
zeroth-order image fell on the S3 (backside-illuminated) chip of the
ACIS-S array.  
The standard data filtering resulted in a
net exposure time of 47.082~ks.
To analyze these archival data, we used CIAO\footnote{See
http://cxc.harvard.edu/ciao} software version 3.0.2 and the {\it
Chandra}\/ calibration database version 2.26.

\section{Results}

\src\ was detected with a total zeroth order count rate of
$1.95\times10^{-2}\ \cts$.  The image was not affected by pile-up.
The centroid position was
R.A. = $19^{\mathrm h}24^{\mathrm m}33\fs076$,
decl. = $+50\arcdeg14\arcmin29\farcs21$ (J2000.0),
which agrees to within
$0\farcs11$
with the {\it Hipparcos}\/ position of the optical counterpart HD~182917
\cite[]{hipparcos}. This offset is well within {\it Chandra}'s 90\%
pointing accuracy\footnote{See http://cxc.harvard.edu/cal/ASPECT/celmon}
of $0\farcs6$.  The binned
 \centerline{\epsfxsize=8.5cm\epsfbox{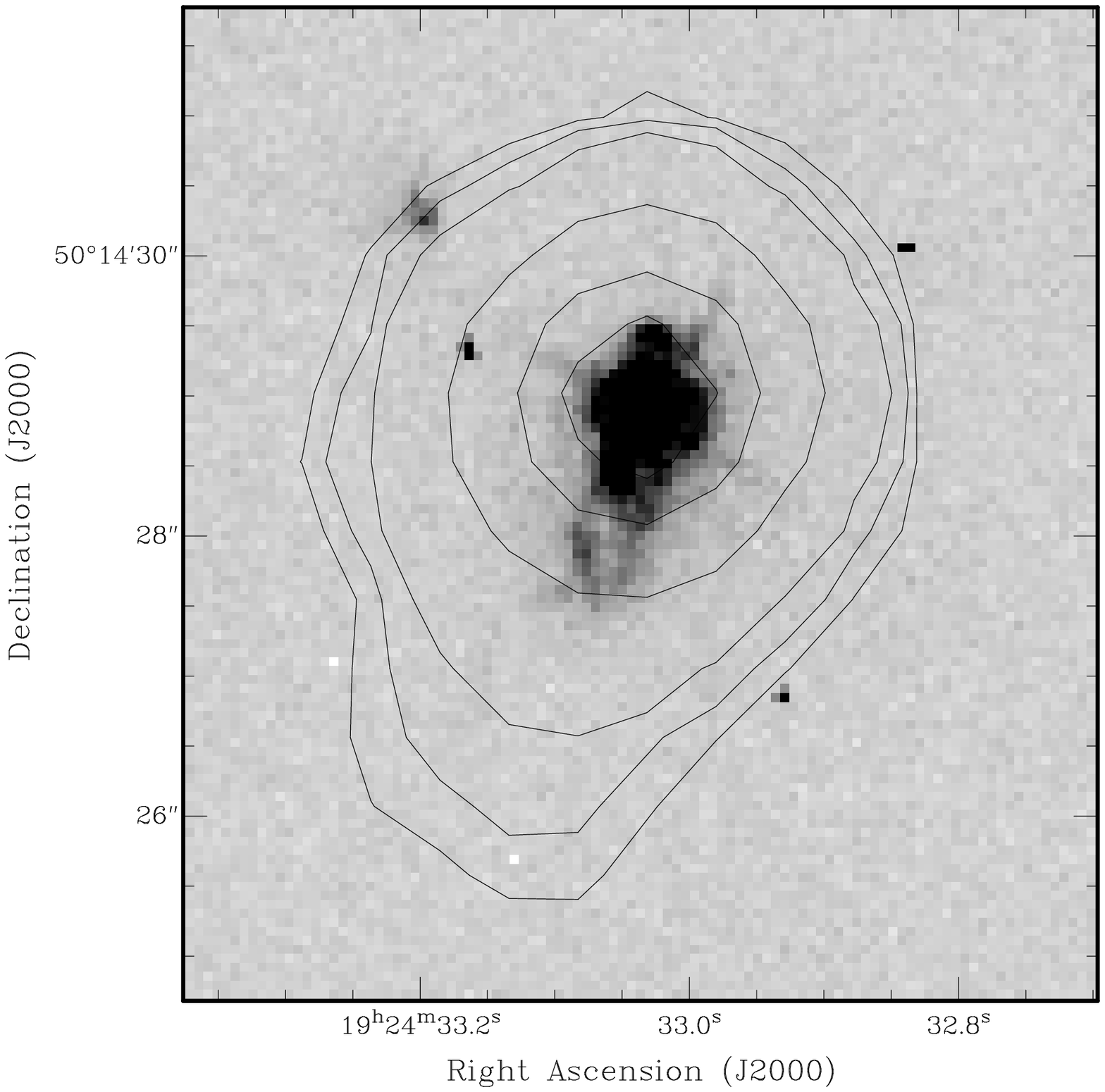}}
 \figcaption[]{ {\it Chandra}\/ \& {\it HST}\/ images of the jet in \src.
The contours show the zeroth order {\it Chandra}\/ image from the 2001
March 27 observation, after smoothing with a Gaussian filter of radius 1".
The contour values are 0.73, 1.2, 2, 10, 30, 50 and 90 counts
per pixel.
The greyscale shows the
{\it HST}\/ WFPC2 image from 1999 August 12 in the O~{\sc III} 495.9~nm \&
500.7~nm filter (i.e. filter F502N); these data are also shown in Fig.
4c of \cite{eyres02}.
 \label{jet} }
\bigskip
\noindent
zeroth order image of \src\ is shown in Fig. \ref{jet}.   

The source exhibited a faint extension $\sim2$--$4\arcsec$ to the
south-southeast.  At the source position, $38\farcs1$ from the optical
axis, no significant deviations from the optimal point-spread function
(PSF) are expected (the 80\% encircled energy radius for an on-axis
point source is $0\farcs685$ at $\approx1$~keV).
At the satellite roll angle used, the readout axis was perpendicular
to the observed extension.
To verify the source extension, we generated a
simulated point-spread function (PSF) with the {\it Chandra}
Ray-Tracer (ChaRT\footnote{See http://cxc.harvard.edu/chart}) in
combination with the Model of AXAF Response to X-rays
(MARX\footnote{See http://space.mit.edu/CXC/MARX}) simulation
software,
using the
zeroth-order spectrum (see below) as input.
We 
measured the radial surface brightness profile of the simulated
zeroth-order image,
and the resulting point-spread
function is shown as a
grey ribbon in Fig.  \ref{radial}. We also measured the observed radial
profile of \src, both in the southern region where
the source is extended, and over the rest of the image excluding
this region (see Fig. \ref{jet}).  
The extension we observed in the binned
zeroth-order image
(Fig. \ref{jet}) is apparent as an excess
above the radial surface brightness profiles of both the simulated
image and the non-southern directions of the observed images.

We extracted 43 photons from
the extended southern region, excluding photons within
$1\farcs72$ of the centroid position.
The probability of observing this many photons, given the extent of the
simulated point source and assuming Poisson statistics,
is $1.67\times10^{-13}$, corresponding to a significance
of $7.4\sigma$. 
The position angle of the extended emission (measured anticlockwise
through east from north) of the principal component of the photon
distribution was $164\arcdeg$, consistent with contemporary
measurements of the \src\ jet from VLA/MERLIN radio and HST optical
measurements \cite[]{crocker02}.  
We therefore conclude that the extension
observed in the {\it Chandra}\/ image is 
due to X-ray emission from the
jet in \src.
The northern optical jet is too small to be resolved with {\it
Chandra}, and we see no evidence for extended emission in that
direction.

The hardness ratio (calculated from the total counts $>2$~keV divided by
the counts $<2$~keV), indicates that the extended emission (hardness ratio
0.65) was significantly softer than the central source (1.69).
The photon distribution was, however, qualitatively similar to that of the
zeroth-order image, with peaks below 1~keV and around 5~keV (Fig.
\ref{spectral}).
To characterize the jet spectrum, we fit the data with an absorbed,
$0.18_{-0.07}^{+0.06}$~keV Raymond-Smith (RS) component for the
emission $<2$~keV and a broad ($\sigma\approx1$~keV) Gaussian at
$5.6_{-0.5}^{+0.6}$~keV ($1\sigma$ errors) for the emission $>2$~keV,
each with a neutral interstellar absorbing column density 
fixed at $7\times10^{20}\ {\rm cm^{-2}}$ \cite[from the H{\sc i}
survey of][]{bellHI}. We found an acceptable fit with a C-statistic
\cite[appropriate for low-count per bin regimes;][]{cash79}
of 4.0, below which fell only 6\%
of 1000 Monte-Carlo simulated spectra. 
Extrapolating this model
beyond the {\it Chandra}\/ energy limits, we estimate an unabsorbed
bolometric flux of $1.2\times10^{-13}\ \epcs$, giving a jet luminosity of
$9\times10^{29}\ \eps$. 

We also fit the total zeroth-order spectrum,
taking photons from within a circle of
radius $\approx5\arcsec$ around the centroid position.  We used
photons from an annulus 5--$10\arcsec$ from the centroid position as
background.
The spectrum exhibited two broad peaks, around 1 and 5 keV, as well as a
narrow 6.64~keV line
(Fig. \ref{spectral}), similar to
the earlier ASCA observation \cite[]{ezuka98}.
 \centerline{\epsfxsize=8.5cm\epsfbox{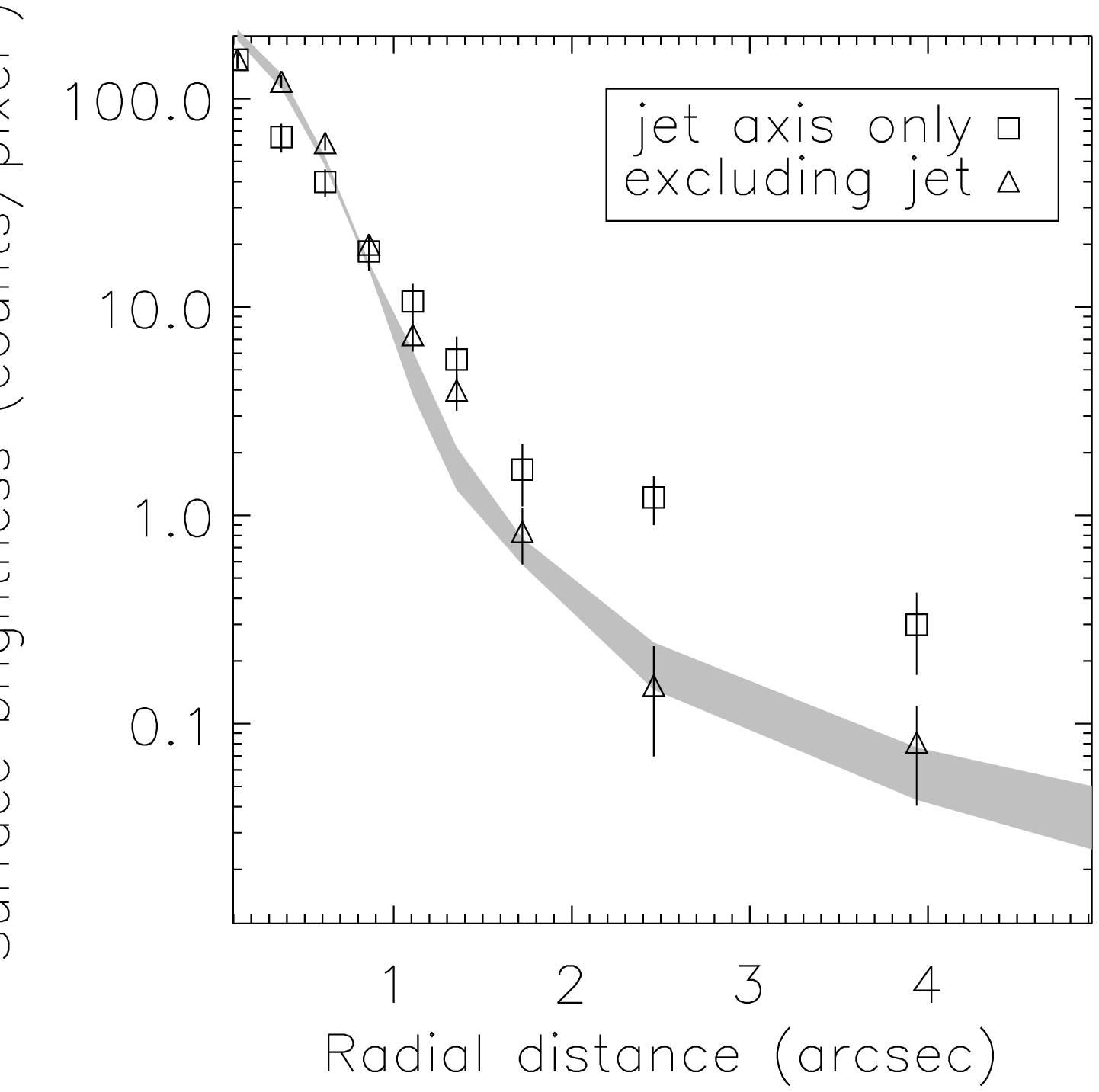}}
 \figcaption[]{Radial surface brightness profiles of the zeroth-order {\it
Chandra}\/ image of \src. The
grey ribbon shows the $\pm1\sigma$ limits on the
simulated PSF from ChaRT/MARX, while the triangles show the radial
profiles calculated using annulae excluding the southern extension. The
squares show the radial profile for the southern extension. Error bars
indicate the $1\sigma$ uncertainties.
 \label{radial} }
 \centerline{\epsfxsize=8.5cm\epsfbox{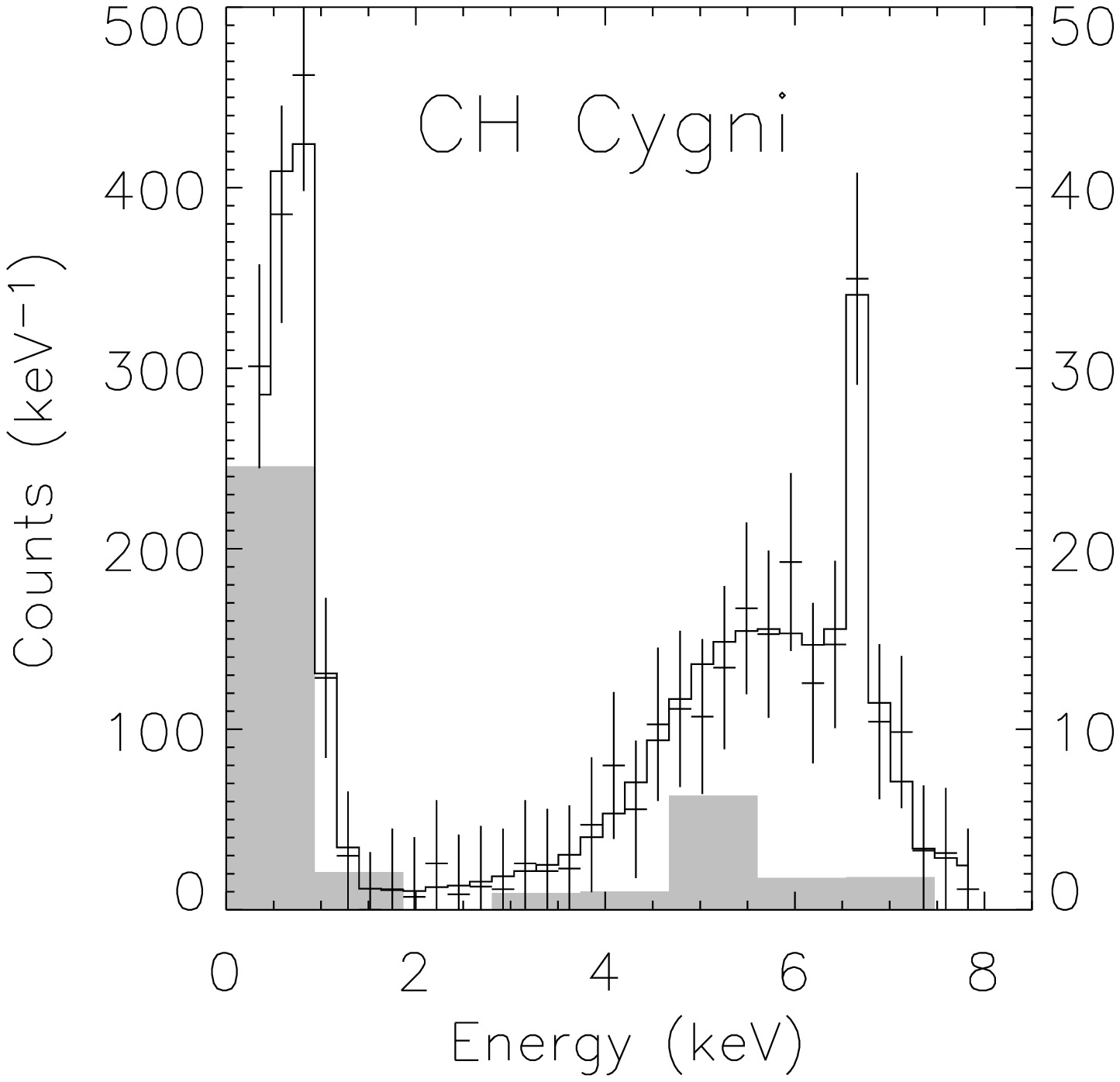}}	
 \figcaption[]{{\it Chandra}\/ spectra from \src.  The zeroth-order
 spectrum is shown as crosses, with error bars indicating the
 $1\sigma$ uncertainties. The dip in emission between 1--4~keV is not
 solely a consequence of a decreased effective area in that region.
 The histogram shows the best-fit model, consisting of 3 Raymond-Smith
 plasma components and a Gaussian to fit the line emission at
 6.64~keV. The grey histgram shows the jet spectrum, scaled by a factor
 of 10 relative to the zeroth-order spectrum (right-hand $y$-axis).
 \label{spectral} }
\medskip
\noindent
Following those authors, we fit the net spectrum with a three-temperature
RS plasma model
plus a Gaussian to 
describe the 
6.64~keV line.
We obtained a C-statistic of 35.2.
53\% of 1000 Monte-Carlo spectral simulations gave a lower value,
indicating a statistically acceptable fit. The spectral parameters were
roughly consistent with the ASCA values.
By extrapolating this plasma-model spectrum beyond the {\it Chandra}\/ energy
limits, we derive an intrinsic (neglecting line-of-sight absorption)
bolometric flux of $1.3\times10^{-11}\ \epcs$, giving a luminosity of
$9.3\times10^{31}\ \eps$ (for a distance of 245~pc).
This luminosity is more than an order of magnitude smaller than typical
values from earlier X-ray measurements of \src\
\cite[e.g.][]{ezuka98}.

\section{Discussion}
\label{sec3}

\subsection{The X-ray Jet in CH~Cygni}

The extended X-ray emission 
in \src\ can be naturally explained by the collision of a collimated
outflow of material with surrounding nebular gas.  Outflow velocities
for \src\ have been estimated 
as 600--$2000\ {\rm km\,s^{-1}}$ and $1100\ {\rm
km\,s^{-1}}$ in 1984/1985
(\citealt{taylor86} and references therein, and \citealt{crocker01}
respectively),
and $>1100\ {\rm km\,s^{-1}}$ in 1998-2000 \cite[]{eyres02}.
For a typical outflow velocity of $\sim1300\ {\rm km\,s^{-1}}$, the
Rankine-Hugoniot jump conditions give
$v_s = (4/3) v_{\rm flow} \approx 1700\ {\rm km\,s^{-1}}$ (where $v_s$
is the shock velocity and $v_{\rm flow}$ is the flow speed in the jet)
for a strong shock.  The temperature behind the shock, $T_{\rm ps}$,
would then be 
\begin{equation}
T_{\rm ps} = \frac{3}{16} \frac{\mu
m_p}{ k_B} v_s^2 = 4 \times 10^7
\left(\frac{v_s}{1700\ {\rm km}\, {\rm
s}^{-1}}\right)^2\left(\frac{\mu}{0.6}\right)\ {\rm K},  
\end{equation}
where $\mu$ is the mean molecular weight, $k_B$ is Boltzmann's
constant, $m_P$ is the proton mass, and we have assumed that the gas
is ideal. 
A plasma with a post-shock temperature of $4 \times 10^7$~K produces
X-rays in the 4-5 keV range;
although we have insufficient jet X-ray photons to perform formal
spectral fitting, we detect X-ray-jet photons at these energies.
The data are thus consistent with material heated in a
strong shock.

Softer X-ray photons are also detected from the jet.  
The time scale for
hot plasma to adiabatically cool by flowing 
out from behind the shock
is given roughly by the width of the jet divided by the sound speed.
For the jet parameters of \src, this time is a few
months.  Thus, 
the lower-temperature jet emission (photons around 1 keV) could be due
to material that has expanded perpendicular to the flow of the jet.
Alternatively, a range of temperatures could arise from multiple shocks in
the jet.

Adopting $v_s = 1700 {\rm km\,s^{-1}}$ and a jet size of $4\arcsec$ on
2001 March 27, we can estimate the epoch of jet production.
For $d=245$~pc, we derive a mean proper
motion for the jet of $1\farcs46\ {\rm yr^{-1}}$.  The extended $Chandra$
emission is therefore consistent with having been produced by material
that was ejected in mid-1998.
At that time, \src\ underwent a transition into an optical high state \cite[see
e.g.][]{crocker01,eyres02,sk03b}.  The detection of extended radio
emission in 1995 by \cite{crocker01} suggests that a 
 \centerline{\epsfxsize=8.5cm\epsfbox{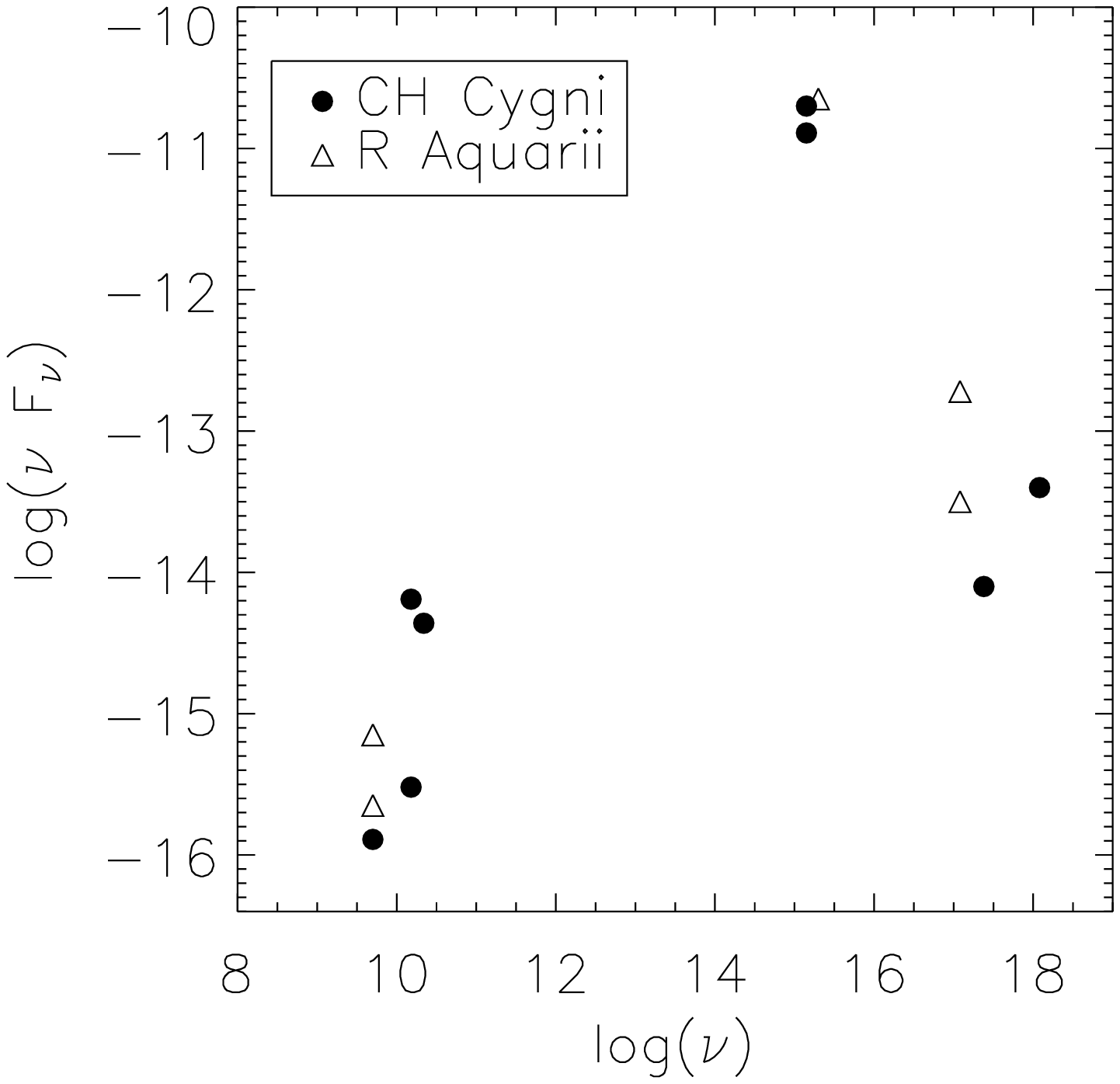}}
 \figcaption[]{Comparison of
 the approximate spectral energy distributions for the jet emission from
 \src\ (filled circles) and R~Aqr (open triangles). The radio data for
 \src\ are from \cite{crocker01} and \cite{kcm98}, and for R~Aqr are
 from \cite{dougherty95}, \cite{lj92} and \cite{kafatos89}. The UV
 data for \src\ are from \cite{eyres02}, and for R~Aqr are from
 \cite{cf03}.  The X-ray measurements for \src\ are from this work,
 and for R~Aqr are from \cite{kellogg01}.
The UV fluxes for \src\
are not extinction corrected, and it is not known whether the UV flux
for R~Aqr from \cite{cf03} is extinction corrected.
The \src\ jet X-ray fluxes are absorption
corrected using the Galactic column only.
\label{sed} }
\bigskip
\noindent
jet ejection also
occurred in association with a transition to the optical high state in
1992.  The other two jet 
ejections, in 1984/85 \cite[]{taylor86} and 1997
\cite[]{kcm98}, were associated with transitions from high
to low optical states.
Thus there appear to be at least two different scenarios in which jets can be
produced in \src.  This situation is reminiscent of X-ray binaries, which
can produce discrete ejections in association with X-ray state changes,
possibly in response to an extreme physical change in the accretion flow
\cite[]{fen00}.

\subsection{Non-relativistic X-ray Jets}

The only other symbiotic star currently known to have an X-ray jet
is R~Aqr \cite[see][]{kellogg01}.  Compared to \src\,
the R~Aqr jet is both more extended on the sky ($30\arcsec$ for the R
Aqr southern jet compared to $4\arcsec$ for \src) and physically
larger (roughly 6000~AU compared to roughly 1000~AU for \src).
Although the measured jet fluxes for the two sources are comparable,
\citet[]{kellogg01} found that few of the jet X-ray photons from R~Aqr
had energies greater than 1 keV.  The R~Aqr jet spectrum revealed
emission lines at 400 and 550 eV and was generally consistent with a
shocked plasma out of ionization equilibrium (the plasma temperatures
for the northeastern and southwestern jets were found to be $2 \times
10^7$ K and $2 \times 10^6$ K respectively).  A follow-up {\it Chandra}\/
observation of R~Aqr revealed that the northern R~Aqr X-ray jet is
expanding at a velocity of $\approx600\ {\rm km\,s^{-1}}$
(E. Kellogg, private communication).  \cite{cf03} compiled and modeled
the full spectral energy distribution (SED) of R~Aqr, and found
that the jet emission is due to a combination of photoionization and
shock heating.  In units of $\nu F_\nu$, where $\nu$ is radiation
frequency and $F_\nu$ is flux density, the R~Aqr jet emission is
strongest in the UV, weakest in the radio, and in between
for the X-rays.  In Fig. \ref{sed} we plot both the SED data compiled
for R~Aqr by \cite{cf03} and that for \src\ obtained from the
literature and our work in the X-rays.
The jet SEDs for R~Aqr and \src\
are quite similar, despite 
R~Aqr appearing to have a slower shock speed, slightly lower
post-shock temperatures, and a larger size.  This similarity suggests
that other symbiotic-star jets might also be expected to emit X-rays.

In addition to symbiotic-star jets, two other types of
non-relativistic jet have been detected in the X-rays: protostellar
jets (in Herbig-Haro objects) and collimated outflows from the central
stars or binaries in planetary nebulae (PNe).  In both cases, the
X-rays also appear to be due to shock heating as the outflowing
material collides with surrounding gas.  
In the two known X-ray-bright protostellar outflows (HH2,
\citealt{pravdo01}; and YL1551~IRS5 = HH~154, \citealt{favata02}), the jet X-ray
luminosities are a few times $10^{29}$ ergs (similar to \src\ and R
Aqr) and the jet plasma temperatures
are $\sim10^6$ K.  \cite{sokat03} summarized six
observations of PNe by {\it Chandra}\/ and {\it XMM-Newton}, four of which revealed
diffuse X-ray emission with temperatures of a few time $10^6$ K.
These four PNe all show evidence for high-velocity, collimated flows
\cite[]{kastner03}. \cite{sokat03}
suggest that
a hidden binary companion within each of
these PNe accrete from the AGB-star wind and produce an X-ray
jet.  They note that the X-ray production mechanism would thus be
similar to that in the protostellar outflows mentioned above, and the
symbiotic star R~Aqr.  In fact, an additional PNe with X-ray-jet
emission, Menzel 3, is known to contain a symbiotic pair at its center
\cite[]{kastner03,zl02,schmeja01}.
The links between the X-ray jets in symbiotic stars, protostellar
outflows, and PNe are thus quite compelling.

\acknowledgments
We thank J. Lazendic for help generating and overlaying the images, as
well as J. Raymond and S. Kenyon for useful discussion.  We also
thank the referee, M. Bode, for his helpful suggestions. This work was
funded in part by the NASA Long Term Space Astrophysics program under
grant NAG 5-9184 (PI: Chakrabarty).  JLS is supported by an NSF
Astronomy and Astrophysics Postdoctoral Fellowship under award
AST-0302055.  This research has made use of data obtained through the
High Energy Astrophysics Science Archive Research Center Online
Service, provided by the NASA/Goddard Space Flight Center.


\clearpage

\end{document}